\documentclass[Afour,sageh,times]{sagej}

\usepackage{moreverb,url}
\usepackage{epsfig} 
\usepackage{graphicx}
\usepackage[caption=false]{subfig}
\usepackage{amsmath}
\usepackage{float}
\usepackage{commath}
\usepackage{epstopdf}
\usepackage[superscript]{cite}
\usepackage{placeins}
\makeatletter
\renewcommand\@biblabel[1]{#1.}
\makeatother
\usepackage[colorlinks,bookmarksopen,bookmarksnumbered,citecolor=black,urlcolor=black]{hyperref}

\newcommand\BibTeX{{\rmfamily B\kern-.05em \textsc{i\kern-.025em b}\kern-.08em
T\kern-.1667em\lower.7ex\hbox{E}\kern-.125emX}}

\def\volumeyear{2020}

\begin{document}

\runninghead{Panda and Warrior}

\title{Mechanics of drag reduction of an axisymmetric body of revolution with shallow dimples}
\author{J P PANDA, J HANDIQUE, H V WARRIOR} 

\affiliation{Department of Ocean Engineering and Naval Architecture\\Indian Institute of Technology, Kharagpur, India}

\corrauth{J P Panda, IIT Kharagpur, INDIA}

\email{jppanda@iitkgp.ac.in}

\begin{abstract}
In this article, the mechanics of drag reduction on an axisymmetric body of revolution by shallow dimples is presented by using the high-fidelity Reynolds Stress Modeling based simulations. Experimental results of drag evolution from published literature at different Reynolds numbers are used to validate the model predictions. The numerical predictions show good agreement with the experimental results. It is observed that the drag of the body is reduced by a maximum of $31\%$ with such shape modification (for the depth to diameter ratio of $7.5\%$ and coverage ratio of $52.8\%$). This arises due to the reduced level of turbulence, flow stabilization and suppression of flow separation in the boundary layer of the body. From the analysis of turbulence states in the anisotopic invariant map (AIM) for the case of the dimpled body, we show that the turbulence reaches an axisymmetric limit in the layers close to the surface of the body. There is also a reduced misalignment between the mean flow direction and principal axis of the Reynolds stress tensor, which results in such drag reduction. The dimple depth to diameter and coverage ratio are also varied to evaluate its effect on drag evolution. 
\end{abstract}


\maketitle

\section{Introduction}
The drag on an axisymmetric body arises either from the pressure of the fluid or from the fluid friction acting along its wetted surface. The major portion of the drag that acts in the direction opposite to the motion of the body is generated from the region that is very close to the boundary of the body where the flow is always turbulent. A detailed understanding of the turbulent flow, its control, the drag evolution and the mechanism of reduction of viscous drag along a body of engineering interest will lead to technological advancements in field of engineering design.   

Drag reduction by flow control along axisymmetric bodies such as submarines and other underwater vehicles\citep{panda2020review} can lead to optimized energy consumption and reduce the cost of operation \citep{frohnapfel2012money}. The viscous drag amounts to about $90\%$ of the total drag of underwater bodies as reported in Frohnapfel et al.\citep{frohnapfel2007experimental}. For aircrafts the drag can be controlled by delaying the laminar-turbulent boundary-layer transition. Sustained laminar flow over aircraft's wings, nacelles and tail surfaces can result in drag reduction. The methods of control of laminar-turbulent boundary layer transition include application of suction near the leading edge, minimization of leading edge surface roughness, reduction of leading edge sweep or by utilizing periodic discrete roughness elements in the spanwise direction\citep{saric2019experiments}.  There are various flow control methods available by which drag over an axisymmetric body can be reduced\cite{choi2008control}. Frohnapfel et al. \citep{frohnapfel2007experimental} used surface embedded grooves for drag reduction. They observed that turbulence reaches an axisymmetric state over the grooved wall and the turbulence fluctuations are highly suppressed in the boundary layer of the grooved wall, which resulted in significant drag reduction. Krieger et al.\citep{krieger2018toward} designed an anti-turbulence surface for maximum drag reduction in channel flows and observed a drag reduction of 60 percent lower than the flat surface. The main purpose of such an anti-turbulence surface is to damp the turbulence stresses in the near-wall flow layers. Watkins et al.\citep{watkins2017flow} used bio-inspired coatings to reduce drag of an AUV\citep{panda2020review,mitra2020experimental}. They mainly have used micro-fibres to modify the surface of the AUV and to suppress flow separation. Toonder et al.\citep{den1997drag} used polymer additives\citep{xi2019turbulent} in turbulent pipe flow for reduction of drag. Itoh et al.\citep{itoh2006turbulent} used seal fur surface in a rectangular channel to check the drag reducing capability of the surface and observed a drag reduction of 12 percent. Degroot et al.\citep{degroot2016drag} studied the drag reduction mechanics of a turbulent channel flow using stream-wise grooves.\citep{abderrahaman2017analysis} used an anisotropically permeable surface\citep{gomez2018turbulent} for drag reduction. With increase in stream-wise permeability the drag was observed to be reduced. However, in their analysis it was noticed after certain value of permeability the drag reduction failed. The drag of an axisymmetric body can also be minimized by dimples. Such research started from the aerodynamic studies on golf-balls. The golf-ball dimples can reduce the drag coefficient of a sphere by 50 percent\citep{choi2006mechanism}. The drag coefficient can be defined as, $C_D=F_D/(0.5\rho U_0^2 A)$ , here $F_D$ is the drag force, $U_0$ is the free stream velocity, d is the diameter of the sphere, $\rho$ is the density of the fluid and A is the cross-sectional area of the sphere.

There are numerous studies available in literature in which experimental and numerical studies are performed to study the flow characteristics over dimpled bodies\citep{tay2014development} such as cylinders\citep{bearman1993control}, spheres\citep{choi2006mechanism} and for internal flow cases such as channels\citep{tay2015mechanics,lienhart2008drag,won2005comparisons,ligrani2001flow}. The effect of dimpling of a sphere on drag reduction was investigated in Choi et al.\citep{choi2006mechanism}, where they observed the drag of the sphere was reduced as much as $50\%$, in comparison to the non-dimpled sphere. They also noticed that the reduced drag coefficient was nearly same for a certain range of Reynolds numbers. Tay et al.\citep{tay2015mechanics} observed that shallow dimples in a channel increase the stream-wise vorticity in the flow field. They considered two depth to diameter ratios of the dimples for their simulations. For a lower range of Reynolds numbers up to 35000, they observed a maximum drag reduction of 3 percent.   

There are no such studies found in literature in which, the detailed flow dynamics and corresponding drag evolution over axisymmetric bodies of revolution\citep{kumar2018large,posa2016numerical,posa2020numerical,mitra2019effects,jimenez2010intermediate,ashok2015asymmetries,posa2018large,bridges2003experimental} such as submarine models with dimpled surface are analyzed. In this study, the detailed dynamics of turbulent flow evolution and its damping over an axisymmetric dimpled body is presented. Experimental results from published literature\citep{jagadeesh2009experimental} for the case of the non-dimpled body are used to validate the numerical model predictions. A series of numerical simulations based on high fidelity Reynolds stress models are performed for analyzing the effect of variation of depth to diameter ratio and coverage area to total surface area ratios on drag evolution along the axisymmetric body. The turbulence structure in the form of Reynolds stress tensor are presented along the body and finally anisotropy invariant map is used to demarcate the state of turbulence for the smooth and dimpled body.   
\begin{figure}
\captionsetup[subfigure]{justification=centering}
\centering
\subfloat[]{\includegraphics[height=1.03cm]{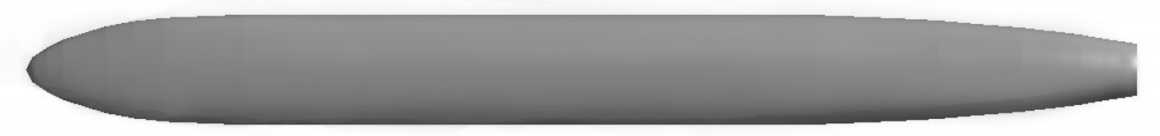}}\\
\subfloat[]{\includegraphics[height=3.2cm]{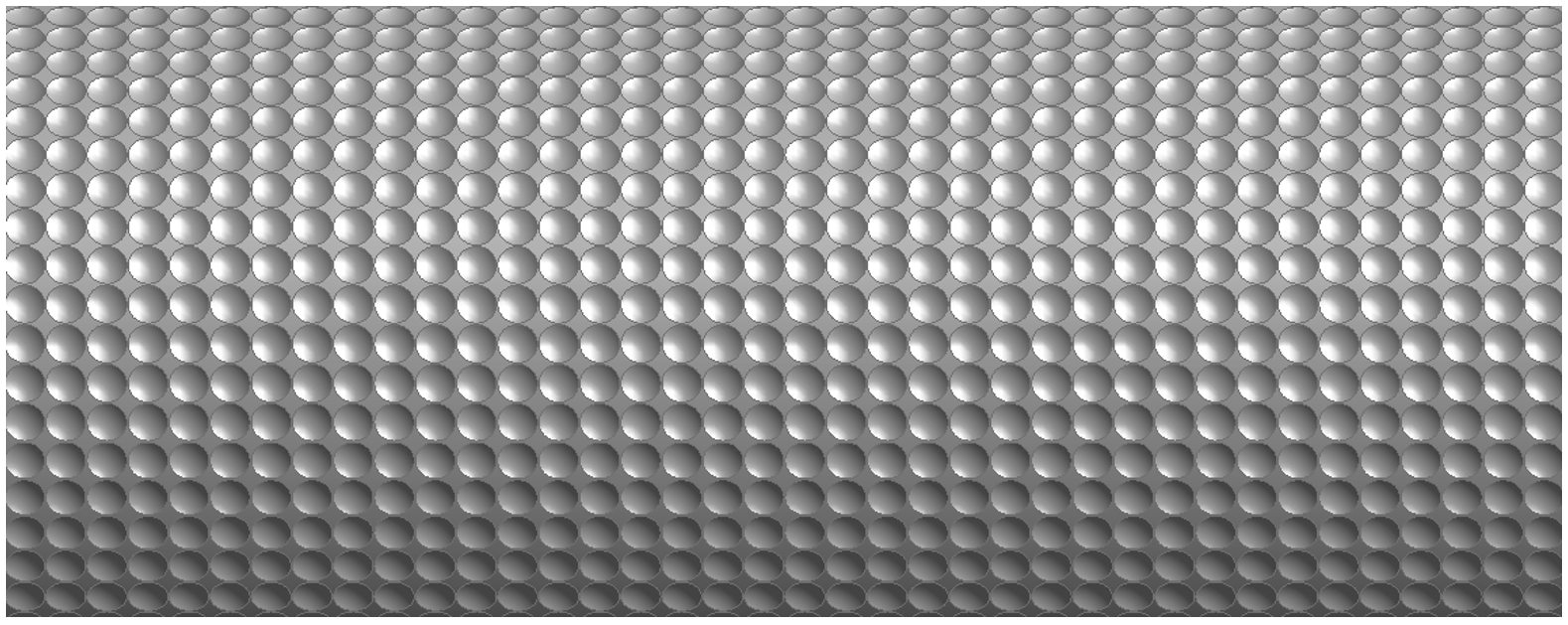}}
\caption{a) Geometry of the axisymmetric body of revolution(non-dimpled), which has similar shape as of DARPA SUBOFF model\citep{groves1989geometric}  developed by \citep{huang1978stern, groves1989geometric} b) The dimpled body with magnified mid-section.\label{fig:1}}
\end{figure}
\section{Numerical modeling}
The present numerical setup consists of a model of an axisymmetric underwater body Fig.\ref{fig:1}. Basically, the model is a predecessor of the DARPA suboff submarine model\citep{huang1978stern,groves1989geometric}. The length and maximum diameter of the model are 1.4 and 0.14 meters respectively. The detailed dimensions of the model are available in\citep{huang1978stern}. The mid-section of the body was dimpled over a length of 0.7m (starting from 0.25m to 0.95 meter from the beginning towards the end) to check its effect on corresponding drag reduction. The surface diameter of the dimple is taken as 14mm. The depth of the dimple is varied by changing the depth to surface diameter ratios from 2.5 to 7.5 percent. The depth to surface diameter ratios are not increased further since that can adversely affect the strength of the body. The coverage ratio of the dimples were also varied to check its effect on drag evolution. In the numerical analysis x, y and z correspond to stream-wise, vertical and span-wise directions respectively. The velocities are u, v and w in the x, y and z directions respectively. 

For the numerical simulations, a key criterion is the selection of the turbulence model. Turbulence models are simplified closures that seek to relate complex, high order quantities such as the Reynolds stress, in terms of lower-order quantities like the mean rate of strain, etc. There are many different turbulence modeling approaches available, such as 1-equation models, eddy viscosity based models, Reynolds stress models, etc and numerous models within each approach. The use of eddy viscosity based models, such as the $k-\epsilon$ and $k-\omega$ models is the norm in industrial simulations of turbulent flows, due to their lower computational expense, computational stability and robustness. 
In this investigation, we utilize the Reynolds Stress Models (RSM)\citep{speziale1991modelling}  to predict the turbulence damping and corresponding drag reduction over the dimpled body. The central difference between the approach of eddy viscosity based models and Reynolds stress closures lies in their computation of the Reynolds stress tensor. Eddy viscosity based approaches use the conceptualization of a turbulent viscosity to relate the turbulent kinetic energy to the instantaneous  mean rate of strain. On the other hand, Reynolds Stress Models explicitly solve the transport equations for the evolution of individual components of the Reynolds stress tensor \citep{pope2001turbulent}. Due to this eddy viscosity hypothesis, eddy viscosity based models are unable to correctly capture the anisotropy of turbulent flows and thus, replicate the directional effects of turbulence \citep{mompean1996predicting, mishra2019linear, sjogren2000development}. Similarly, they are unable to adhere to physically permissible solutions for many complex flows \citep{mishra2014realizability}. In such cases, Reynolds Stress Models offer better performance as they explicitly compute the transport of the Reynolds stress tensor components. However, a key assumption of eddy viscosity based turbulence models is that the eigen-directions of the Reynolds stress tensor are perfectly aligned with those of the mean rate of strain \citep{mishra2015hydrodynamic}. While this alignment is not true for most cases of complex turbulent flow, this misalignment is even more pronounced for regions where flow separation and reattachment occur. As a result for turbulent flows with flow separation and reattachment eddy viscosity models often give unsatisfactory results \citep{craft1996development}. Similarly eddy viscosity models assume that the Reynolds stress tensor forms an adequate basis to describe the state of the turbulent flow field\citep{mishra2016sensitivity}. This assumptions leads to discrepancies in the model predictions for complex turbulent flows of engineering interest\citep{mishra2019estimating}. In investigations, the use of eddy viscosity based models for turbulent flow separation problems has led to substantial errors in the prediction of the location of flow separation, the location of flow reattachment and the size of the separation bubble. These inaccuracies have cascading effects on the model's predictions of the forces on the body, including the drag forces on the body. Such flow separation and reattachment are a primary focus of our study as they have a seminal effect of the drag over a body. In this light we utilize the Reynolds Stress Modeling approach in our investigation.

In summary, RSM based models\citep{panda2020reliable,warrior2014improved} have higher potential to replicate the complex flow physics\citep{manceau2014investigation} over the dimpled body in comparison to the eddy viscosity models\citep{pope2001turbulent}(eddy viscosity models are known to be unreliable for flows with separation, strong pressure gradient and curvature), since RSM does not rely on simplifying assumptions in defining the Reynolds stress in term of the local flow parameters, rather directly employ equations for the Reynolds stress components\citep{panda2017improved,mishra2013intercomponent} and one scale determining equation in the flow field and to predict the flow evolution mechanism and by using such models the complete structure of turbulence in terms of the Reynolds stress components can be obtained directly\citep{mishra2017toward,panda2019review}. The predictive capability of such models for flow along cylinders is validated in \citep{monte2011analysis}, which shows the predictions are quite satisfactory for a wide range of Reynolds numbers. In contrast to direct numerical simulations and large-eddy simulations\citep{posa2020numerical,kumar2018large} (for LES, Kumar and Mahesh\citep{kumar2018large} have used 8192 processors to simulate the flow past an axisymmetric body of revolution) the computational cost is very less. 

The modeled transport equation for the Reynolds stress can be written as\citep{panda2018representation,mishra2010pressure}:
\begin{equation}
\begin{split}
&\partial_{t} \overline{u_iu_j}+U_k \frac{\partial \overline{u_iu_j}}{\partial x_k}=P_{ij}-\frac{\partial T_{ijk}}{\partial x_k}-\eta_{ij}+\phi_{ij},\\
&\mbox{where},\\ 
& P_{ij}=-\overline{u_ku_j}\frac{\partial U_i}{\partial x_k}-\overline{u_iu_k}\frac{\partial U_j}{\partial x_k},\\
&\ T_{kij}=\overline{u_iu_ju_k}-\nu \frac{\partial \overline{u_iu_j}}{\partial{x_k}}+\delta_{jk}\overline{ u_i \frac{p}{\rho}}+\delta_{ik}\overline{ u_j \frac{p}{\rho}},\\
&\eta_{ij}=-2\nu\overline{\frac{\partial u_i}{\partial x_k}\frac{\partial u_j}{\partial x_k}}  \\
&\phi_{ij}= \overline{\frac{p}{\rho}(\frac{\partial u_i}{\partial x_j}+\frac{\partial u_j}{\partial x_i})}\\
\end{split}
\end{equation}

\begin{figure}
\centering
\includegraphics[width=0.48\textwidth]{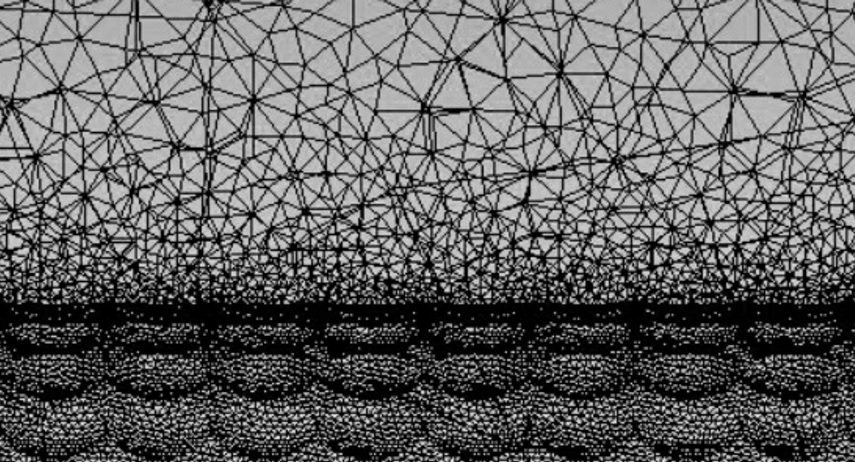}
\caption{Mesh in local region around the dimpled AUV.\label{fig:23456}}
\end{figure}

In equation 1, $P_{ij}$ denotes the turbulence production, the diffusive transport is denoted as $T_{ijk}$ , $\epsilon_{ij}$: dissipation rate tensor and $\phi_{ij}$: pressure strain correlation. The complex flow physics resulting from flow over dimpled surfaces can be accurately captured by a standard pressure strain correlation model. For the present simulations, the linear model of \citep{speziale1991modelling} is used, in which the pressure strain correlation has the form:
\begin{equation}
\begin{split}
& \phi_{ij}^{(R)}=(C_1-C_1^*II^{0.5})K S_{ij}+\\ & C_2K(b_{ik} S_{jk}+b_{jk} S_{ik}-2/3b_{mn} S_{mn}\delta_{ij})\\ & +C_3K(b_{ik} W_{jk}+b_{jk} W_{ik})
\end{split}
\end{equation}
Following standard practice, the closure coefficients are taken as  $C_{1}=0.8$, $C_1^{*}=1.3$, $C_2=1.25$ and $C_3=0.4$, as is outlined in the original study \cite{speziale1991modelling}.

In order to study the effect of the dimpling of the axisymmetric body, we performed numerical simulations of the physical problem under consideration. A rectangular three dimensional solution domain is constructed over the body. The governing equations are solved by using a finite volume method\citep{anderson1995computational}. The no-slip and no-penetration boundary conditions are applied at the walls. A line-by-line tridiagonal matrix element algorithm (TDMA) is used for solving the discretized system of linear algebraic equations. A fully implicit scheme is used for the time integration. The semi-implicit method for pressure linked equations (SIMPLE) is used for the pressure velocity coupling. A more densely spaced grid is used close to the wall of the axisymmetric body as shown in Fig.\ref{fig:23456}. We had used unstructured mesh in our simulations. A mesh sensitivity study is conducted for four different meshes with successive refinement leading to an increase in the total number of cells in the flow domain. For the meshes utilized the total number of cells in the domain were 0.9 Million,1.8 Million, 2.5 Million and 3.5 Million, respectively. We find that the quantities of interest calculated using the final two levels of mesh refinement do not show appreciable changes and consequently, we utilize the mesh with 2.5 Million elements in the results of this investigation. 

\begin{figure}
\includegraphics[width=0.5\textwidth]{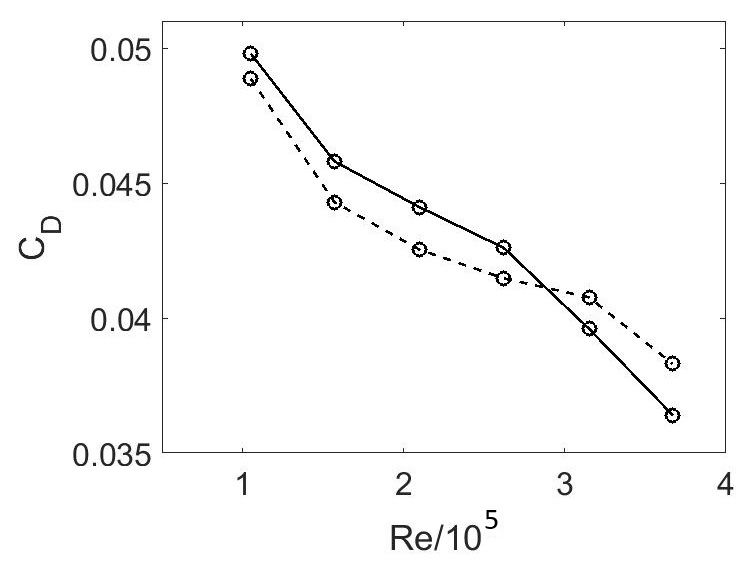}
\caption{Variation of drag coefficient with Reynolds numbers, the dashed and solid lines correspond to the RSM predictions and the experimental results of\citep{jagadeesh2009experimental} respectively.\label{fig:2}}
\end{figure}

While using wall functions with Reynolds Stress Models, the best practice guidelines\citep{fluent2015ansys, salim2010wall, ariff2009wall} for simulations outline that the $y+$ value should be in the range 30$< y+ <$300. Since the Reynolds stress model was used for the numerical simulations with the wall functions, the values of y+ for the four meshes utilized in the mesh independence study were calculated to be 190, 130, 70 and 40 respectively. The corresponding first layer thickness for the three different meshes was calculated accordingly. For y+ values 190,130, 70 and 40, the drag coefficient values are 0.0315, 0.0326, 0.0382 and 0.0383 respectively, for the Reynolds number $3.67\times10^5$, (the drag coefficient from the experimental data for the same conditions was 0.0389). The drag coefficient values predicted with the mesh of 2.5 Million cells matches well with experimental results of\citep{jagadeesh2009experimental} for all six Reynolds numbers (ranging from $1.05\times{10^5}$ to $3.75\times{10^5}$). The RSM predictions of drag coefficients are presented for different Reynolds numbers with solid lines in fig. \ref{fig:2}. Experimental results of \citep{jagadeesh2009experimental} are shown in dotted lines. The numerical predictions are in good agreement with the experimental results. In all the numerical simulations the solution has been iterated till convergence at each time step.

\section{Results and discussion}
\subsection{Effect of dimpling of the body on drag reduction}
\begin{figure}
\includegraphics[width=0.45\textwidth]{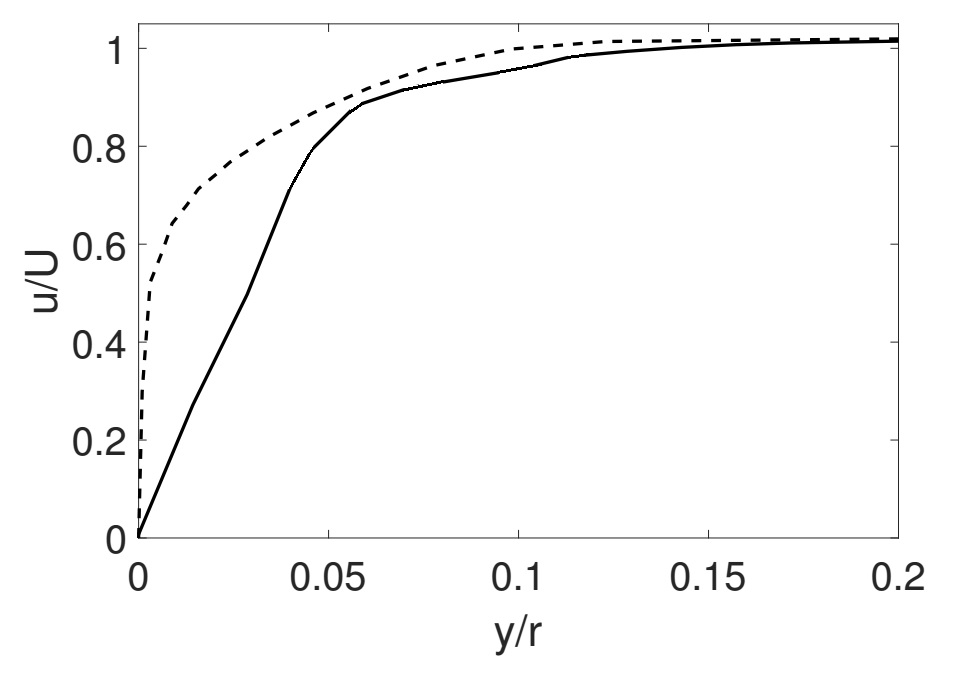}\\
\includegraphics[width=0.45\textwidth]{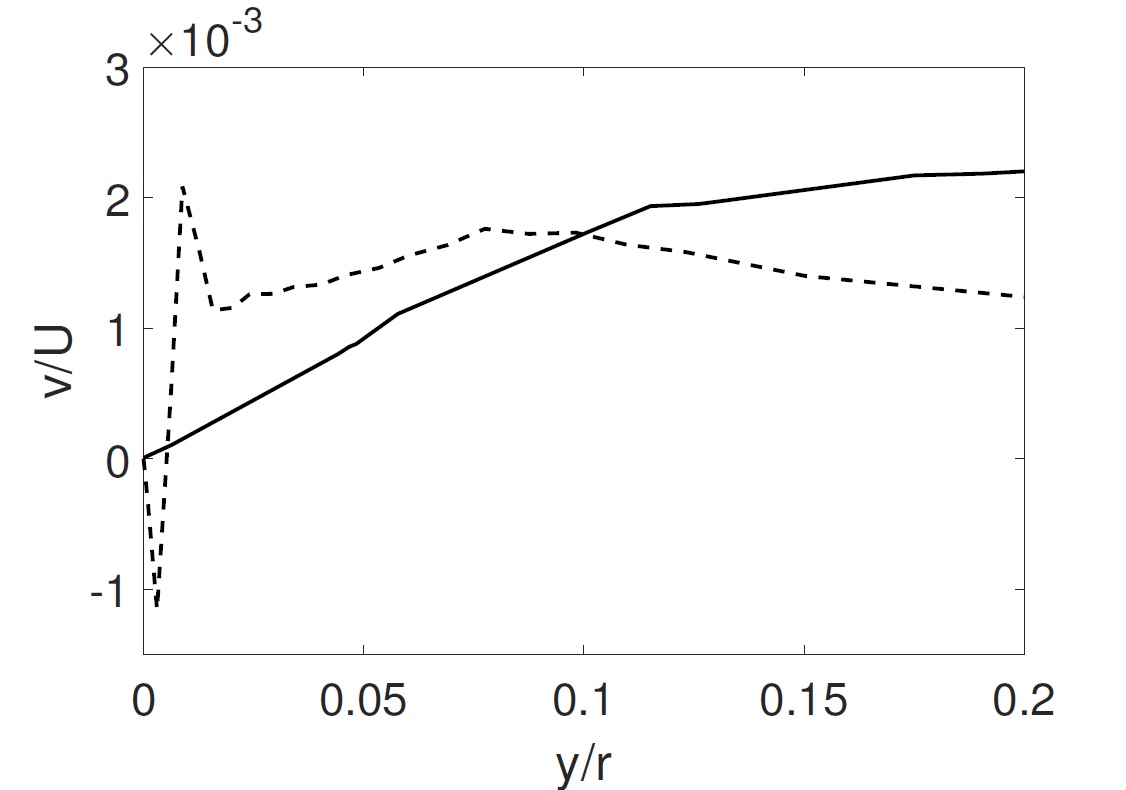}\\
\caption{Comparison of the velocity distribution for dimpled (solid line) and non-dimpled body (dashed line). The data were taken at x/r=8.5 (at the beginning of the body x/r=0).\label{fig:3}}
\end{figure}

 Fig. \ref{fig:3}a shows the variation of velocity in the vertical direction over the body. The velocity was non-dimensionalized with respect to the corresponding free stream velocity. From the reduced slope of velocity profile over the dimpled body, it is clear that flow is stabilization is manifested as observed in the investigation of Posa and Balaras\citep{posa2020numerical}. The magnitude of x-velocity is decreasing, because of the interaction of eddies generated by dimples with the longitudinal vortices in the streamwise direction. The larger slope of the velocity profile for the non-dimpled body signifies the rapid production of turbulence in the boundary layer. An analysis of velocity component in the radial(wall normal) direction (Fig. \ref{fig:3}b) reveals that, for the case of non-dimpled body in the regions very close to the wall, the wall normal velocity is negative, which signifies the existence of flow separation. However, for the dimpled body the wall normal velocity is positive, this give insights into the mechanism by which the dimpled body suppress the flow separation.  
 
\begin{figure}
 \includegraphics[width=0.5\textwidth]{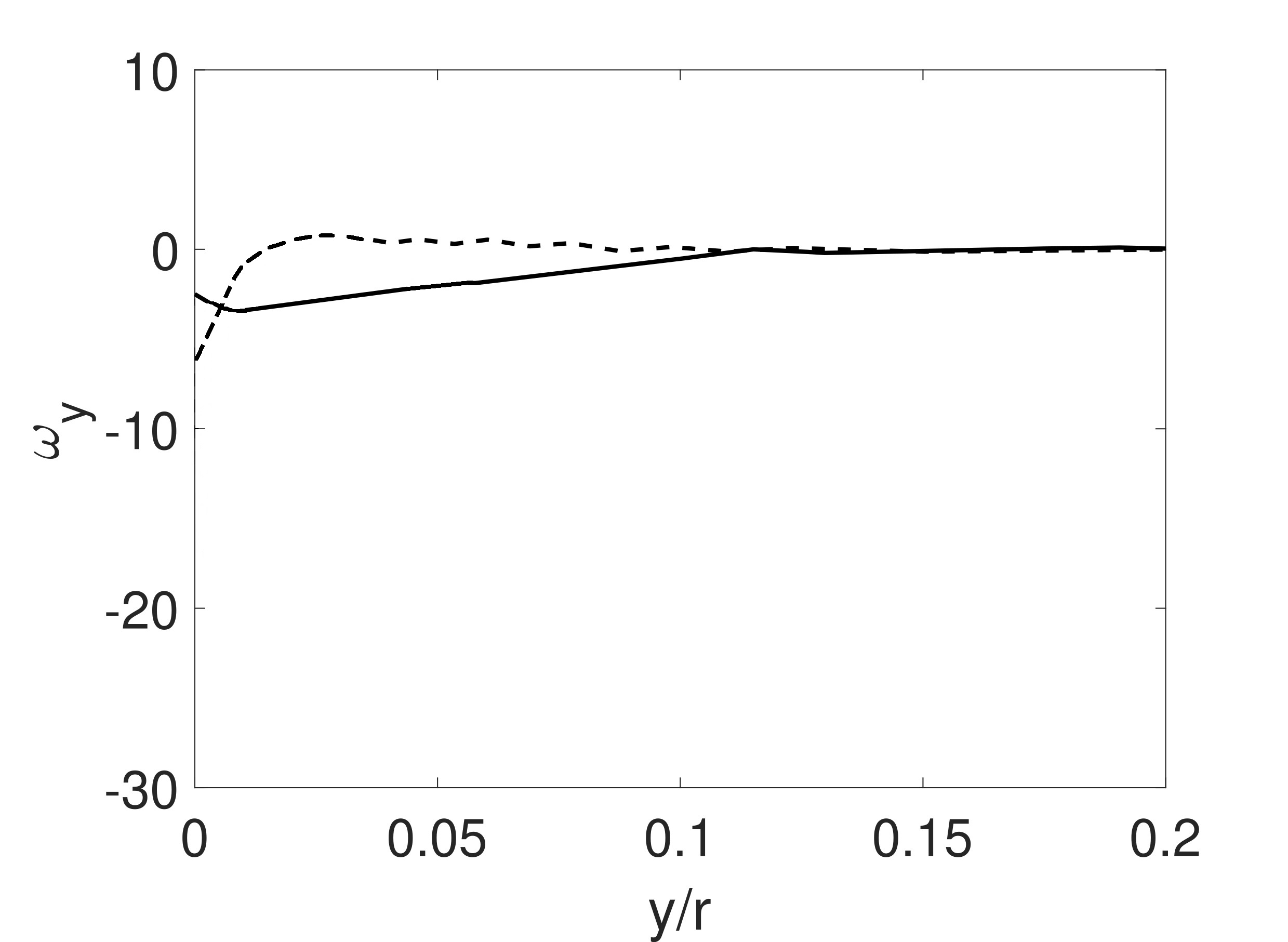}
 \caption{\label{fig:epsart}Comparison of the wall normal vorticity (along the y-direction) distribution for dimpled (solid line) and non-dimpled (dashed line) body.\label{fig:3333}}
 \end{figure}

 In the turbulent boundary layer, the existence of wall-normal vertical fluid structures (eddies) are the main source of skin friction drag as reported in\citep{fukagata2002contribution,robinson1991coherent}. These are associated with a larger magnitude of wall-normal vorticity and the higher vorticity is generated by the interaction of the large eddies with the wall of the axisymmetric body. In figure \ref{fig:3333} the variation of wall-normal vorticity is presented. The solid and dashed line corresponds to the vorticity distribution for the dimpled and non-dimpled bodies respectively. There is a reduction of the wall-normal vorticity for the dimpled body since the dimples act as large eddy break-up (LEBU) devices\citep{choi1994active}, which breaks the larger turbulence generating eddies into smaller dissipative eddies in layers very close to the body resulting in a net reduction of turbulent energy of eddies for interaction with the wall. For the dimpled body, as shown in figure \ref{fig:3333} the wall-normal vorticity is mainly reduced in the layers very close to the wall of the body, which signifies that the near-wall coherent structures has a larger effect on the skin friction drag evolution along the body. Dimples mainly provide suction effect and dampen the coherent structures\citep{park1999effects}.     

\begin{figure}
\includegraphics[width=0.5\textwidth]{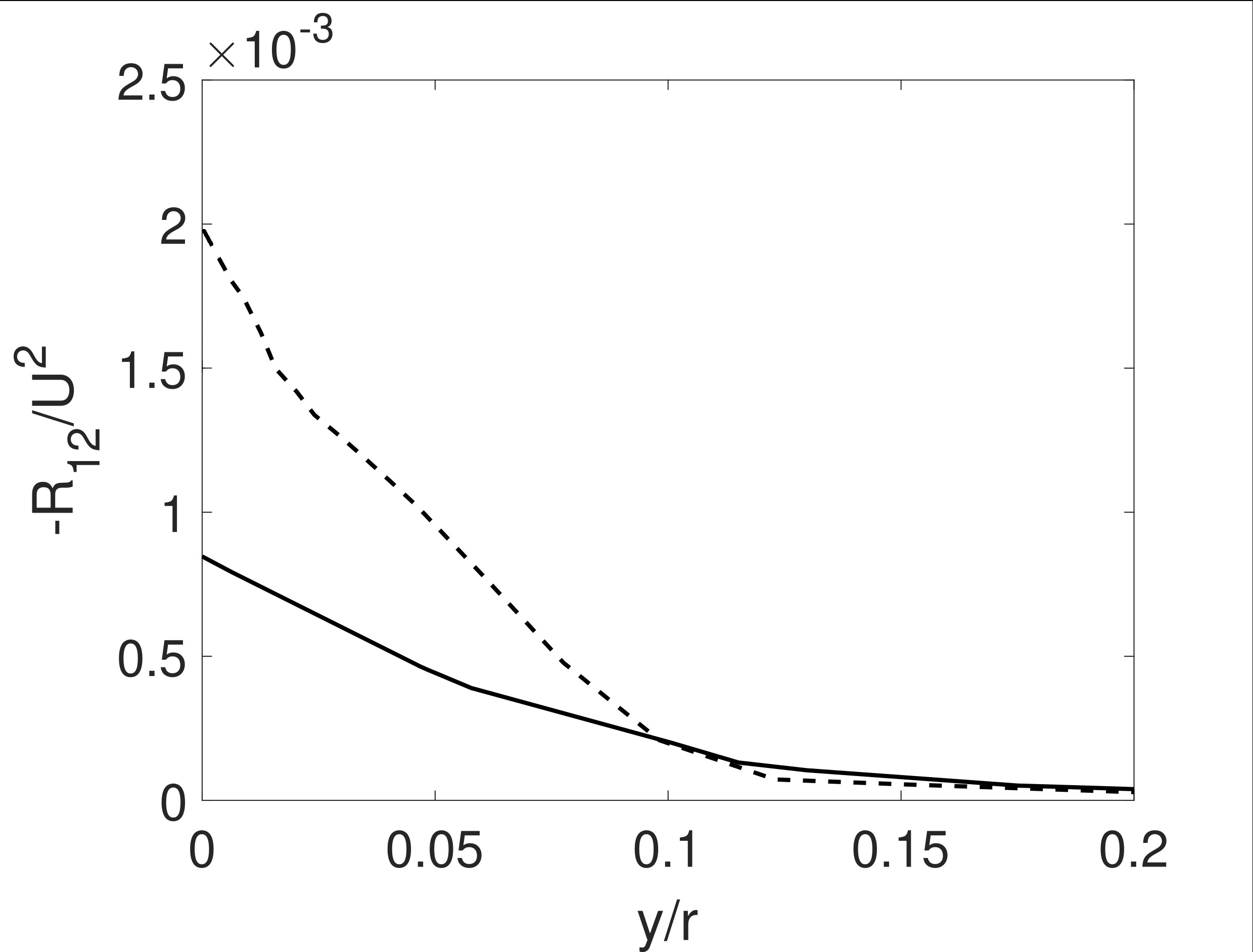}
\caption{The distribution of Reynolds shear stress in the boundary layer of the axisymmetric body. The solid and dashed line correspond to dimpled and non-dimpled bodies respectively.\label{fig:4}}
\end{figure}

 The distribution of turbulence/Reynold shear stress  is shown in Fig. \ref{fig:4}. From the figure it is clear that there is a sharp decrease in the turbulence shear stress, because of the flow damping and stabilization over the dimpled surface. The dimple array mainly suppresses the formation and interaction of wall-normal eddies by generating stream-wise vorticity into the flow field\citep{tay2015mechanics}. 
 
 The turbulence shear stress $R_{12}$ has a direct relationship with the skin friction evolution along axisymmetric bodies\cite{monte2011analysis}:
\begin{equation}
\begin{split}
& C_{f}^T=\frac{1}{f(a)}\int_{a}^{a+1} (1+a-y)(-R_{12})ydy
\end{split}
\end{equation}
where a is the radius of the axisymmetric body and $f(a)=a/2(11/24-a/6)$ is a shape factor. In the formulation of the turbulent skin friction $C_f^T$, The Reynolds stress is weighted by $1+a-y$, which signifies that the near wall turbulence structures have a larger contribution on the skin friction evolution. The reduction in turbulence stresses resulted in skin friction drag reduction, and the total drag of the body is reduced\citep{choi1994active}.

\begin{figure}
\includegraphics[width=0.5\textwidth]{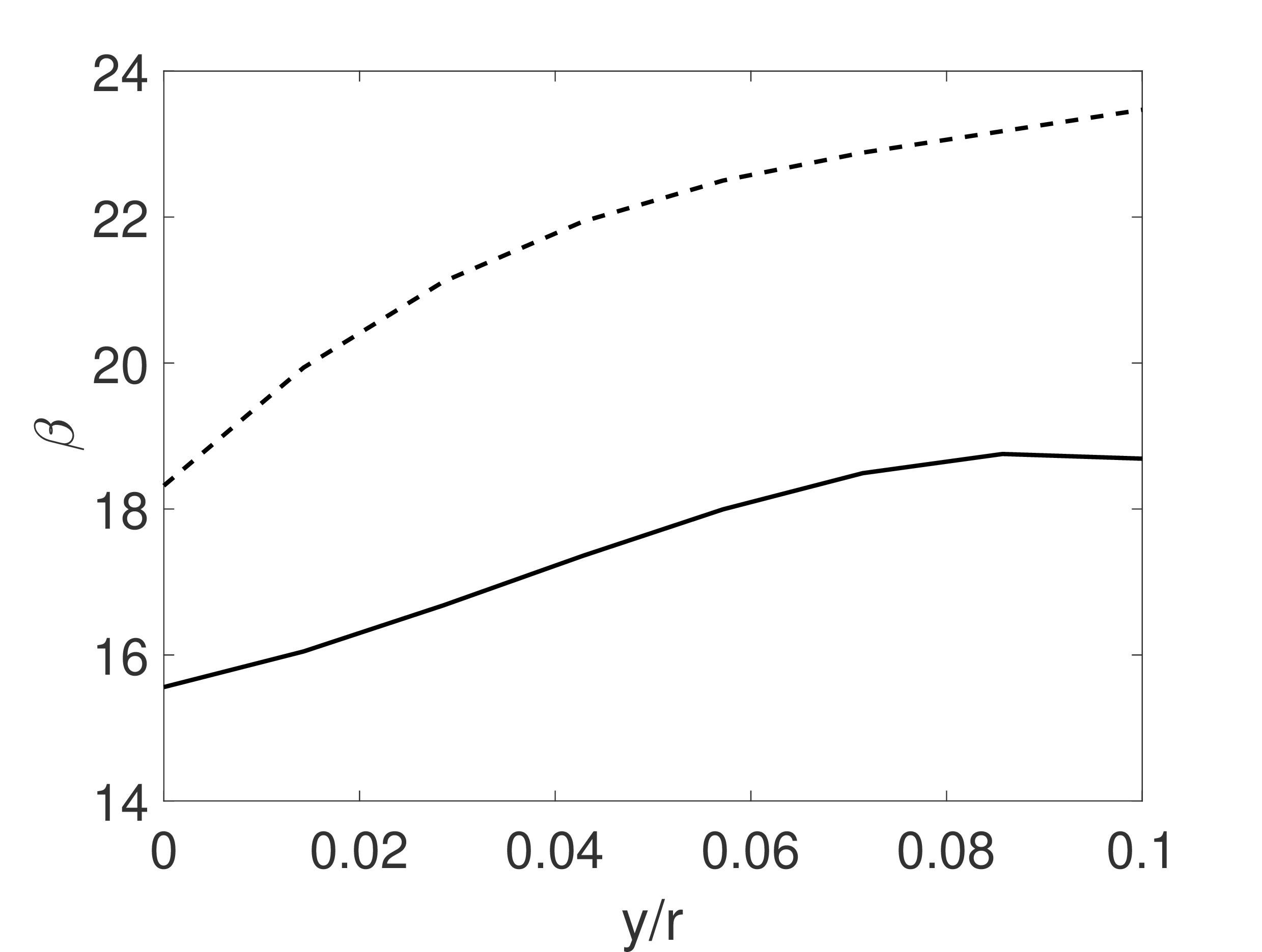}
\caption{The variation of angle $\beta$ in the boundary layer of the body. The solid and dashed line correspond to dimpled and non-dimpled bodies respectively.\label{fig:444}}
\end{figure}

The Reynolds shear stress can be decomposed in terms of the Reynolds stress anisotropy tensor\citep{monte2011analysis}. When considering the rotation of an angle $\beta$ such as $b_{ij}$ is diagonal, $D_{kl}$,
\begin{align*}
& b_{ij}=p_{ik}D_{kl}P_{lj},where\,P=\begin{pmatrix}
cos\beta & sin\beta & 0\\
sin\beta & cos\beta & 0\\
0 & 0 & 1
\end{pmatrix}
\end{align*}
The Reynolds shear stress can be expressed as a function of angle $\beta$, the first two eigenvalues of the Reynolds stress anisotropy tensor and the turbulence kinetic energy\citep{monte2011analysis},

\begin{equation}
\begin{split}
& -R_{12}=-k(\lambda_2-\lambda_1)sin{2\beta}.
\end{split}
\end{equation}
As described in \citep{monte2011analysis}, the reduction of angle $\beta$ has a direct effect on the skin friction drag evolution along the body. The reduction of $\beta$ signifies a reduced misalignment between the mean flow direction and the main axis of the Reynolds stress. As shown in figure \ref{fig:444} there is a sharp reduction in the angle $\beta$ with y/r in the boundary layer of the dimpled body. 




\begin{figure}
\includegraphics[width=0.5\textwidth]{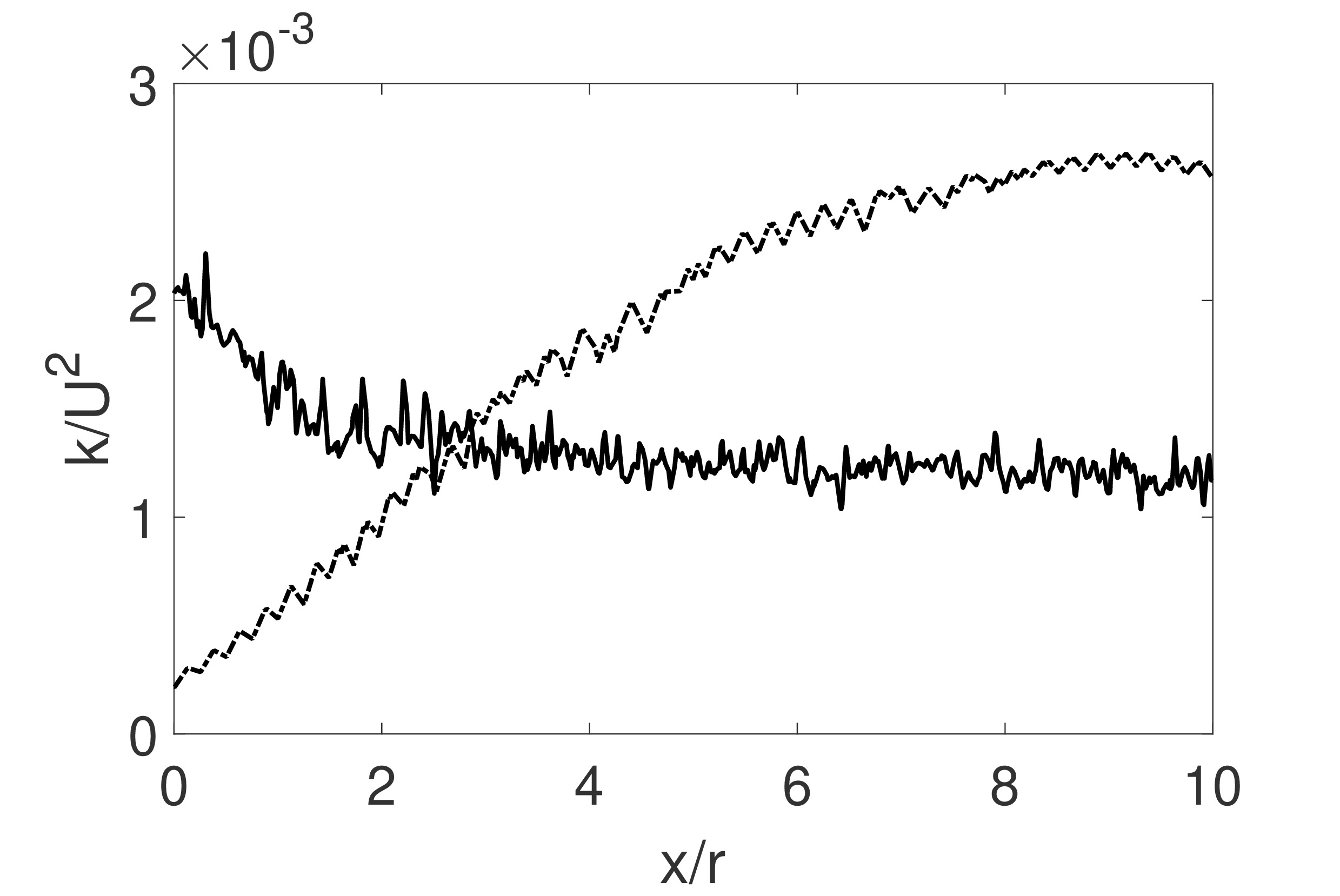}
\caption{ Evolution of turbulence kinetic energy along the length of the body, for y/r=0.07. The solid and dashed-dot line correspond to dimpled and non-dimpled bodies respectively. A decay of turbulence kinetic energy is noticed along the length of the dimpled body. \label{fig:5}}
\end{figure}


\begin{figure}
\includegraphics[width=0.5\textwidth]{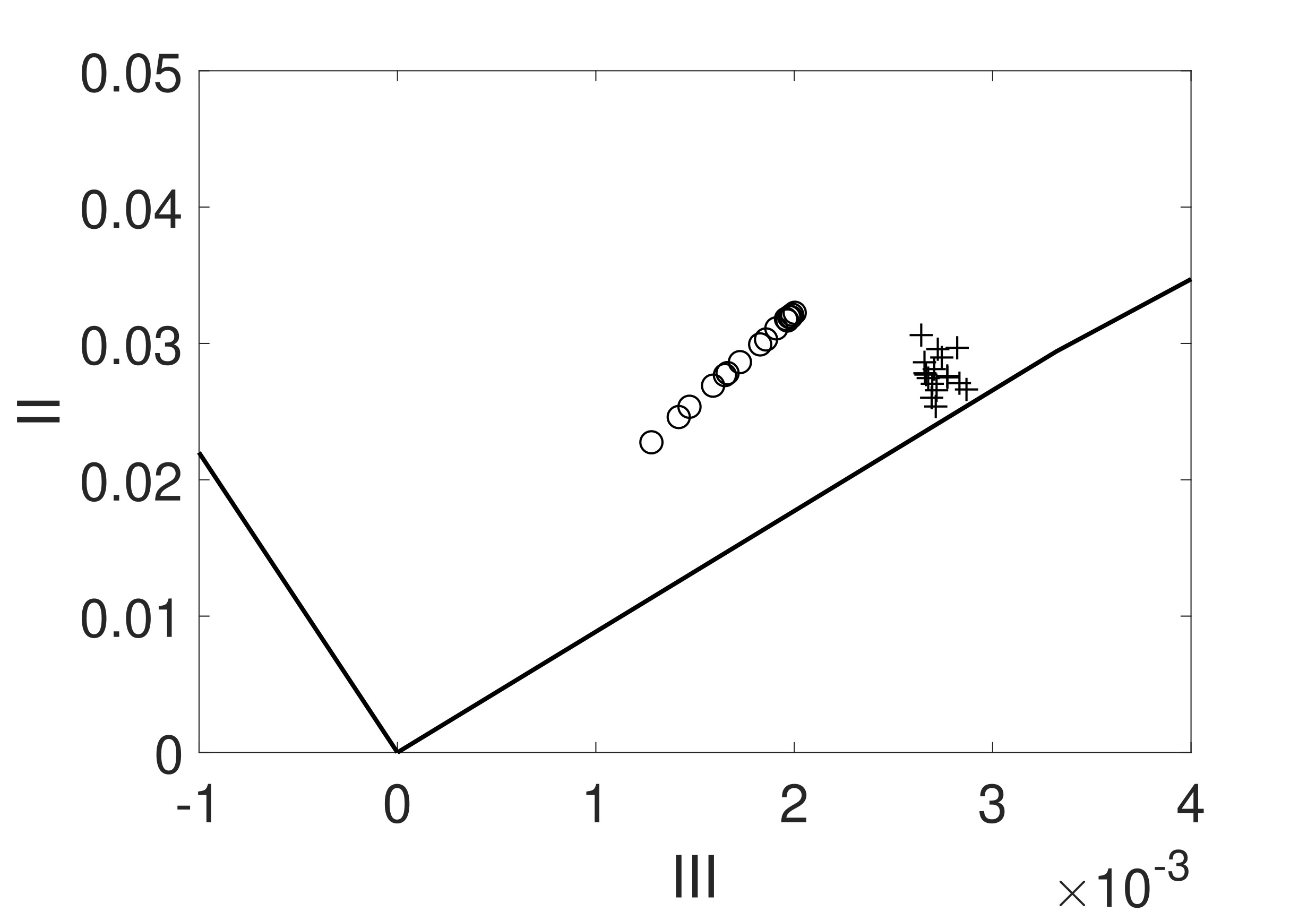}
\caption{The anisotropy invariant map representing different states of turbulence. The circles and plus symbols represent turbulence states over non-dimpled and dimpled body respectively. The data were taken at different stream-wise locations along the body (for y/r=0.07).\label{fig:7}}
\end{figure}
For an accurate  description of turbulence damping over the dimpled body, the variation of turbulence kinetic energy over a horizontal line along the length of the body in the boundary layer (at y/r=0.07) is plotted in fig.\ref{fig:5}. It is noticed that the turbulence along the body is decaying similar to the pattern observed in grid generated turbulence\citep{panda2018experimental}. However, a reverse trend of turbulence evolution is observed over the non-dimpled body, where turbulence is found to be increasing along the length of the body.

Krieger et al.\citep{krieger2018toward} observed that turbulence satisfies statistical axisymmetry in the velocity fluctuations and one component state in the near-wall region, over the anti-turbulence surfaces.
The different states of turbulence (such as one component or two component) along the body can be visualized by anisotropy invariant map (AIM)\citep{lumley1979computational}, with the second $(=-b_{ii}^2/2)$ and third invariants$(=b_{ii}^3/3)$ of the Reynolds stress anisotropy$(b_{ij}=R_{ij}/2k-1/3\delta_{ij})$. $R_{ij}$ is the Reynolds stress tensor. From the AIM paths of turbulence reported in figure \ref{fig:7}, it is observed that the turbulence over the dimpled body is shows a higher degree of axisymmetric character, as all the plus symbols in the AIM figure approaches toward the right side of the AIM where the Reynolds stress ellipsoid corresponds to rod-like axisymmatric states\citep{simonsen2005turbulent}. This is in accordance to the findings of \citep{frohnapfel2007interpretation}, where they noticed that for drag reduced flows the near wall turbulence approaches towards one component limit. However, the turbulence over the non-dimpled body is far from the axisymmetric state.
\subsection{Effect of depth to diameter(s/d) ratio on drag reduction}
\begin{figure}
\includegraphics[width=0.5\textwidth]{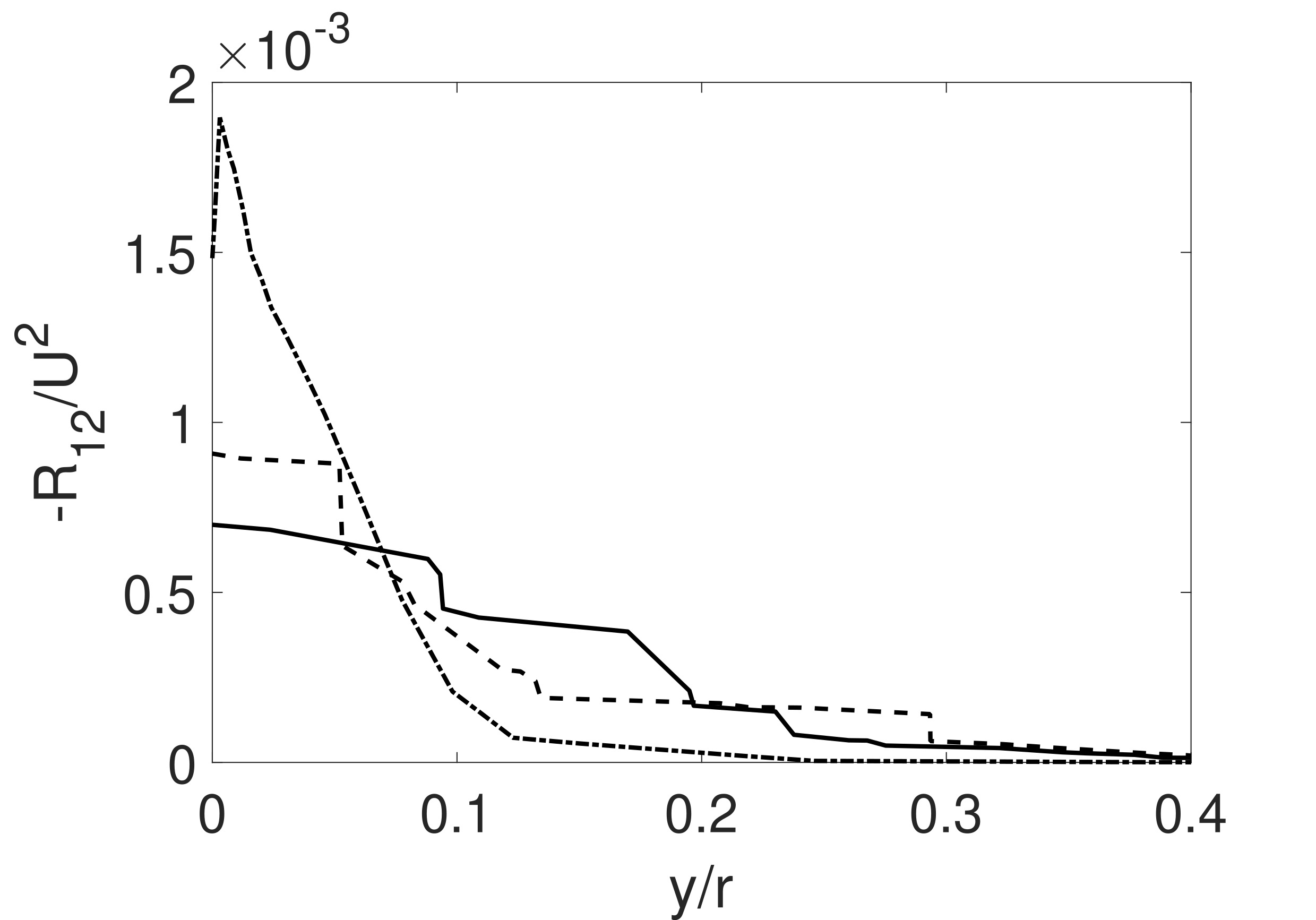}
\caption{The distribution of Reynolds shear stress in the boundary layer of the axisymmetric body. The dashed, dashed-dot and solid lines correspond to 2.5, 5 and 7.5 percent depth to diameter(s/d) ratios respectively. \label{fig:22}}
\end{figure}
\begin{figure}
\includegraphics[width=0.48\textwidth]{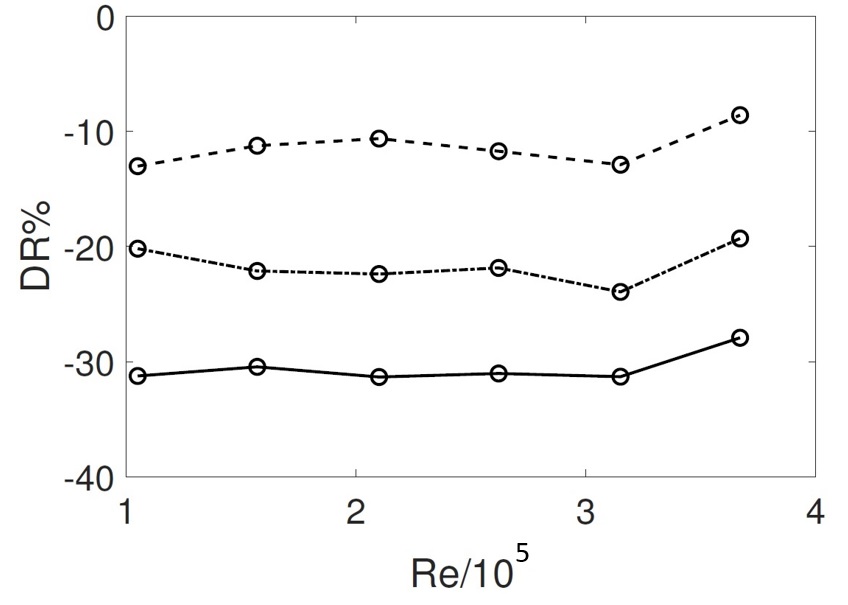}
\caption{Drag reduction achieved with micro-dimples, the dashed, dashed-dot and solid lines correspond to 2.5, 5 and 7.5 percent depth to diameter(s/d) ratios respectively.\label{fig:23}}
\end{figure}

To analyze the effect of the ratio of depth ($s$) to surface diameter ($d$) of the dimple shape on the turbulence evolution along the axisymmetric body of revolution, this s/d ratio is varied from from 2.5 percent to 7.5 percent. This ratio describes the aspect ratio of the dimple and is a characteristic of its geometry. As reported in literature for the case of dimpled channel flow\citep{tay2015mechanics}, further increase in s/d ratio may result in flow separation, that may enhance drag.

The variation of Reynolds shear stress is presented in Fig. \ref{fig:22}. The dashed, dashed-dot and solid lines correspond to the turbulence stresses for s/d ratios 2.5, 5 and 7.5 percents respectively. With increase in s/d ratio from 2.5 to 5 percent, a decrease in turbulence stresses in all three components of the turbulence stresses are noticed, However with further increase in s/d ratio, an increase in Reynolds stress components at upper layers starting from y/r ratios 0.06 to 0.3 is observed. Consequently it is not advisable to increase the depth of the dimple further, which may enhance the turbulence level further and will result in drag increase, which is undesirable. Although, the turbulence level is higher for a range of y/r ratios for s/d equals to 7.5 percent, a over all drag reduction is achieved at all the Reynolds numbers in Fig. \ref{fig:23}, since turbulence level at the layers close to the body is less in comparison to the other two s/d ratios. This depict the fact that, the near wall turbulence structure has larger effect on viscous drag evolution in comparison to the free stream turbulence for the flow along axisymmetric bodies. In Fig. \ref{fig:23} the net drag reduction percentage for different values of depth to diameter ratio of the dimples are presented. The negative sign in the figure signifies that the drag is reducing with the use of the dimpled surface. An increase in drag reduction was noticed with an increase in depth to diameter ratios. Maximum drag reduction was observed for the depth to diameter ratio of $7.5$ percent (31 percent). The depth to diameter was not enhanced further since that can have an adverse effect on the strength of the body. \citep{choi2006mechanism} had observed a drag reduction of 50 percent for dimpled spheres, the drag reduction of the present axisymmetric body can further enhanced by applying dimples over the entire surface area of the body.

\subsection{Effect of coverage ratio (a/A) on drag reduction}
\begin{figure}
\captionsetup[subfigure]{justification=centering}
\centering
\subfloat[]{\includegraphics[width=0.42\textwidth]{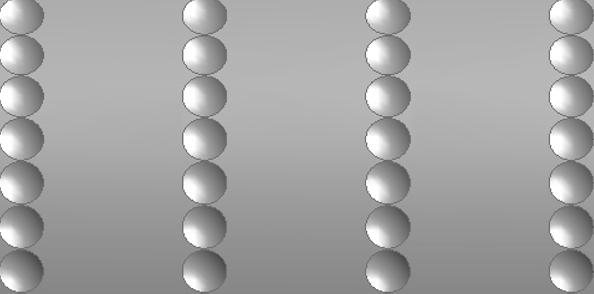}}\\
\subfloat[]{\includegraphics[width=0.42\textwidth]{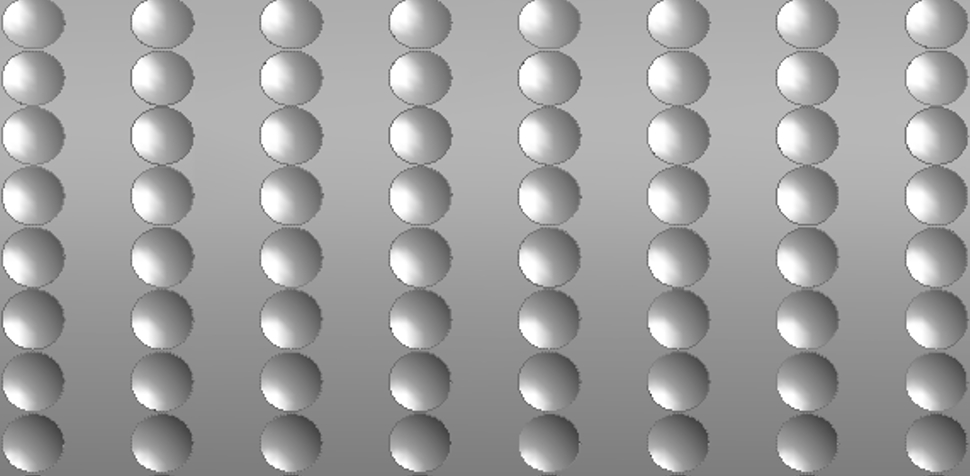}}\\
\subfloat[]{\includegraphics[width=0.42\textwidth]{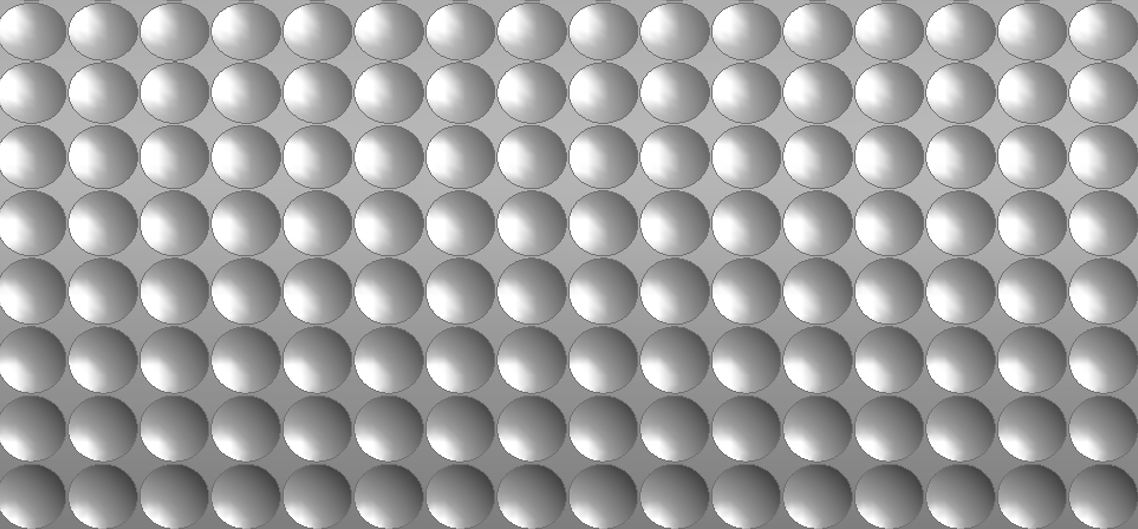}}
\caption{Dimpling of the body for coverage area variation. only a magnified portion is provided. Dimensions not to scale. a) a/A=13.2 percent, b) a/A=26.4 percent, c) a/A=52.8 percent respectively.\label{fig:31}}
\end{figure}
\begin{figure}
\includegraphics[width=0.5\textwidth]{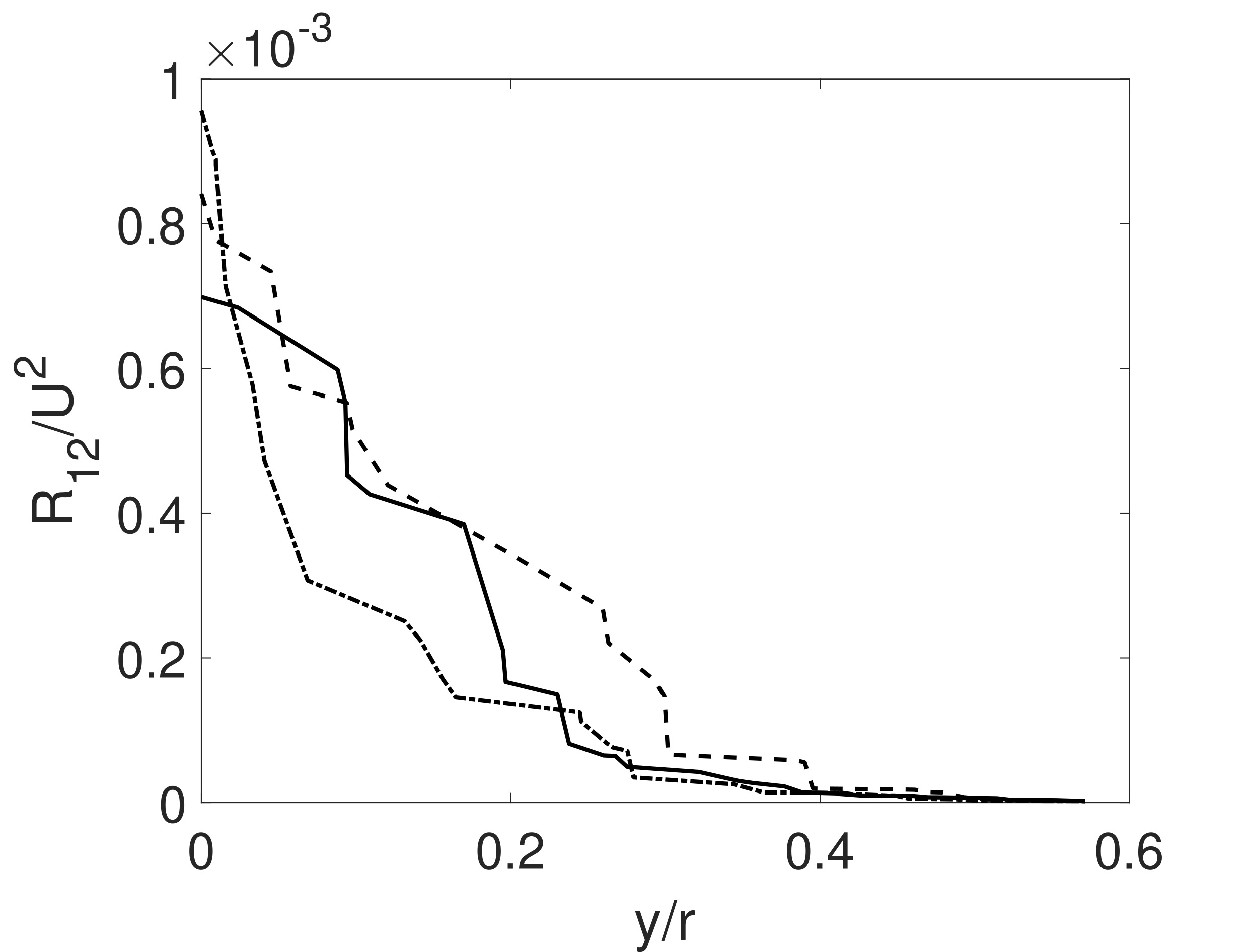}
\caption{The distribution of Reynolds stress components in the boundary layer of the axisymmetric body. The dashed, dashed-dot and solid lines correspond to 13.2, 26.4 and 52.8 a/A ratios respectively.\label{fig:32}}
\end{figure}

\begin{figure}
\includegraphics[width=0.5\textwidth]{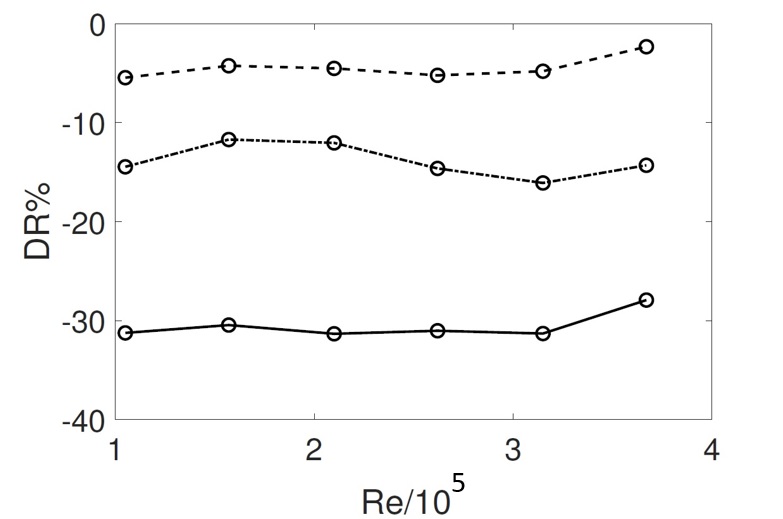}\\
\caption{The drag reduction achieved with different a/A ratio. Dashed, dashed-dot and solid lines correspond to 13.2, 26.4 and 52.8 a/A ratios respectively.\label{fig:33}}
\end{figure}
The coverage ratio(a/A) can be defined as the ratio of area occupied by dimples(a=total number of dimples multiplied by area occupied by one dimple over the body) divided by the total surface area of the body over which dimples were made. Accordingly three different a/A ratios (13.2, 26.4 and 52.8 percent) are considered in the simulations as shown in Fig. \ref{fig:31}. For varying the coverage ratio, the dimple densities were varied only along the length of the body as shown in Fig. \ref{fig:31}. The dimple densities in the circumferential direction were not varied.   

In Fig. \ref{fig:32} the variation of turbulence stresses for different coverage ratios are presented. A similar trend of trend of turbulence evolution (as of s/d variation) is observed for different coverage ratios. With increase in coverage ratio form 13.2 percent to 26.4, all the three components of turbulence stresses near the wall was found to be increasing, however, the stresses at the layers above y/r=0.02 turbulence level was very less. With further increase in a/A ratio a decrease in near wall turbulence was observed. This is clear that both near wall and far wall turbulence contributions has effect on the viscous drag evolution.       


\section{Concluding remarks}

The reduction of drag on bodies is an critical line of research for engineering design as it may lead to  better performance and efficiency. In this article, the mechanics of drag reduction of an axisymmetric body of revolution with shallow dimples have been presented. The effect of dimpling of the body on the evolution of turbulence structure along the body was analyzed. It was observed that the dimples have the ability to suppress the turbulence intensity in the boundary layer of the body, which resulted in a maximum drag reduction of $31$ percent. Additionally the dimples on the surface lead to a change in the componentiality of the flow field evidenced in the state of turbulent flow anisotropy. From the anisotropy invariant map it was observed that due to this modification the state of turbulence componentiality transitions closer to the axisymmetric state, in the boundary of the body and over the dimpled body the turbulence was found to be decaying similar to the pattern observed in decaying grid turbulence. With variation of depth to diameter ratio of dimples, the near wall turbulence level was decreased, so an decrease in drag coefficient was noticed. A similar turbulence and drag evolution pattern was observed  for coverage ratio variations. The s/d and a/A ratio were not increased further, since those may reduce the strength of the body. This is undesirable, since underwater bodies often cruise in marine environments where operating pressure is very high. Here we recommend s/d and a/A ratios 7.5 and 50 percent respectively, for safer operations in Naval Engineering applications. A detailed analysis based on strength of the body will be more appropriate for such recommendations, which can be considered as an future course of work. 

\bibliographystyle{plainnat}
\bibliography{asme2e}
\end{document}